\documentclass[aps,noshowpacs,tightenlines,nobalancelastpage,floatfix,twocolumn]{revtex4}

\usepackage{amssymb}
\usepackage{amsmath}
\usepackage{graphicx}
\usepackage{dcolumn}
\usepackage{bm}

\begin{document}

\title{Generation of Nonclassical Photon Pairs for Scalable Quantum
Communication with Atomic Ensembles}
\date{\today }
\author{A.~Kuzmich, W. P.~Bowen, A. D. Boozer, A. Boca, C.~W. Chou, L.-M.
Duan, and H.~J.~Kimble} \affiliation{Norman Bridge Laboratory of
Physics 12-33, California Institute of Technology, Pasadena, CA
91125}


\maketitle

\textbf{Quantum information science attempts to exploit capabilities from
the quantum realm to accomplish tasks that are otherwise impossible in the
classical domain \cite{qi1}. Although sufficient conditions have been
formulated for the physical resources required to achieve quantum
computation and communication \cite{qcc}, there is an evolving understanding
of the power of quantum measurement combined with conditional evolution for
accomplishing diverse tasks in quantum information science \cite%
{bose99,KLM,br1}. In this regard, a significant advance is the invention of
a protocol by Duan, Lukin, Cirac, and Zoller (DLCZ) \cite{duan01} for the
realization of scalable long distance quantum communication and the
distribution of entanglement over quantum networks. Here, we report the
first enabling step in the realization of the protocol of DLCZ, namely the
observation of quantum correlations for photon pairs generated in the
collective emission from an atomic ensemble. The nonclassical character of
the fields is evidenced by the violation of a Cauchy-Schwarz inequality, }$%
\tilde{g}_{1,2}^{2}(\delta t)\nleq \tilde{g}_{1,1}\tilde{g}_{2,2}$\textbf{,
with }$\tilde{g}_{i,j}$\textbf{\ as normalized correlation functions for the
two fields }$(i,j)=(1,2)$\textbf{. As compared to prior investigations of
nonclassical correlations for photon pairs produced in atomic cascades \cite%
{clauser74} and in parametric down conversion \cite{mandel99}, our
experiment is distinct in that the correlated $(1,2)$ photons are separated
by a programmable time interval }$\delta t$\textbf{, with }$\delta t\simeq
400$\textbf{nsec in our initial experiments.}

The theoretical proposal of \textit{DLCZ} \cite{duan01} is a probabilistic
scheme based upon the entanglement of atomic ensembles via detection events
of single photons in which the sources are intrinsically indistinguishable,
and generates entanglement over long distances via a quantum repeater
architecture \cite{br2}. The \textit{DLCZ} scheme, with built-in quantum
memory and entanglement purification, is well within the reach of current
experiments and accomplishes the same objectives as previous more complex
protocols that require still unattainable capabilities \cite{br2,vanenk98}.

In our experiment, we demonstrate a basic primitive integral to the \textit{%
DLCZ} scheme. Specifically, an initial \textit{write} pulse of (classical)
light is employed to create a state of collective atomic excitation as
heralded by photoelectric detection of a first photon $1$. After a
programmable delay $\delta t$, a subsequent \textit{read} pulse interrogates
the atomic sample, leading to the emission of a second (delayed) photon $2$.
The manifestly quantum (or \textit{nonclassical}) character of the
correlations between the initial \textit{write }photon $1$ and the
subsequent \textit{read} photon $2$ is verified by way of the observed
violation of a Cauchy-Schwarz inequality for coincidence detection of the $%
(1,2)$ fields \cite{clauser74}. Explicitly, we find $[\tilde{g}%
_{1,2}^{2}(\delta t)=(5.45\pm 0.11)]\nleq \lbrack \tilde{g}_{1,1}\tilde{g}%
_{2,2}=(2.97\pm 0.08)]$, where $\tilde{g}_{i,j}$ are normalized correlation
functions for the fields $(i,j)$ and $\delta t=405$ ns is the time
separation between the $(1,2)$ emissions. The capabilities realized in our
experiment provide an important initial step toward the implementation of
the full protocol of DLCZ, which would enable the distribution and storage
of entanglement among atomic ensembles distributed over a quantum network.
Extensions of these capabilities could facilitate scalable long-distance
quantum communication \cite{duan01} and quantum state engineering \cite{qse}%
. For example, by employing spin-polarized samples in optical-dipole or
magnetic traps \cite{metcalf99}, it should be possible to extend the
interval $\delta t$ to times of several seconds.

Our experiment arises within the context of prior work on spin squeezing
\cite{kitagawa91,wineland92}, and in particular on atomic ensembles where
significant progress has been made in the development of methods to exploit
collective enhancement of atom-photon interactions provided by optically
thick atomic samples \cite%
{kuzmich97,kuzmich98,molmer99,hald99,kuzmich00b,julsgaard01}. Instead of
homodyne or heterodyne detection of light as used in spin-squeezing
experiments \cite{hald99,kuzmich00b,julsgaard01}, the \textit{DLCZ} scheme
involves photon counting techniques, which present stringent requirements
for broad bandwidth detection and for the suppression of stray light from
the atomic ensemble.

\begin{figure}[tb]
\includegraphics[width=8.6cm]{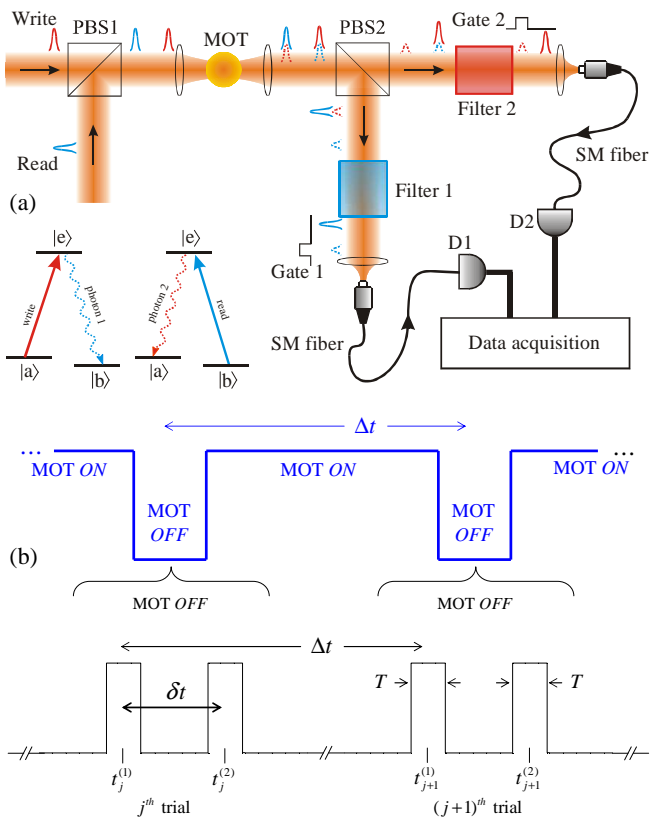}
\caption{A simplified schematic of the experiment is presented, with (a)
providing a diagram of the apparatus and (b) giving the timing sequence for
data acquisition. Further details are as follows. (a) \textit{Write} and
\textit{read} pulses propagate sequentially into a cloud of cold Cs atoms
(MOT), generating pairs of correlated output photons $(1,2)$, with
controlled separation $\protect\delta t$. Fields with frequency near that of
the $|a\rangle \leftrightarrow |e\rangle $ ($|b\rangle \leftrightarrow
|e\rangle $) transition are colored red (blue) here and in the subsequent
two figures. The \textit{write} and \textit{read} pulses have orthogonal
polarizations, are combined into a single input at PBS1 (PBS - polarizing
beam splitter), and are then focussed into the Cs MOT with a waist of
approximately $30\protect\mu $m. The output fields are split by PBS2, which
also serves as the first stage of filtering the (\textit{write}, \textit{read%
}) beams from the $(1,2)$ fields. For example, field $2$ is transmitted by
PBS2 to be subsequently registered by detector \textit{D2}, while the
\textit{read} pulse itself is reflected by $90^{\circ }$ at PBS2 and then
blocked by an acousto-optical modulator that serves as Gate 1. Further
filtering is achieved by passing each of the outputs from PBS2 through
separate frequency filters each of which consists of a glass cell of Cs
vapor optically pumped to place atoms into either $6S_{1/2},F=3$ or $F=4$
\protect\cite{stpetersburg}. The small residual reflected (transmitted)
light of the \textit{write} (\textit{read}) pulse from PBS2 at frequency $%
\protect\omega _{4,4}(\protect\omega _{3,4})$ passes through a
filter cell with atoms in the $F=4(3)$ level. It is thereby
strongly attenuated ($>10^{6} $), while the accompanying
Raman-scattered light as photons $1(2)$ at frequency
$\protect\omega _{3,4}(\protect\omega _{4,4})$ is transmitted with
high efficiency ($\simeq 80$\%). Transmission efficiencies from
the MOT to detectors (\textit{D1}, \textit{D2}) are both about
$30$\% for light with the spatial shape of the \textit{write} and
\textit{read} beams and of the correct polarization. (\textit{D1},
\textit{D2}) have overall quantum efficiencies of approximately
$50$\%\ (photon in to TTL pulse \textit{out}). (b) Gating windows
for the joint detection of photons $(1,2)$ are centered at times
$(t_{j}^{(1)},t_{j}^{(2)})$ for the $j^{\text{th}}$ trial of the
experiment during intervals when the MOT is \textit{OFF}.}
\label{layout}
\end{figure}

As illustrated in Fig. \ref{layout}, an optically thick sample of
three-level atoms in a lambda-configuration is exploited to produce
correlated photons via the following sequence. With atoms initially prepared
in state $|a\rangle $ by optical pumping, a laser pulse\ from the \textit{%
write} beam tuned near the $|a\rangle \rightarrow |e\rangle $ transition
illuminates the sample and induces spontaneous Raman scattering to the
initially empty level $|b\rangle $ via the $|e\rangle \rightarrow |b\rangle $
transition at time $t^{(1)}$. The \textit{write} pulse is made sufficiently
weak so that the probability to scatter one Raman photon into the preferred
forward propagating mode $\psi ^{(1)}(\vec{r},t)$ is much less than unity
for each pulse. Detection of a photon\ in the mode $\psi ^{(1)}(\vec{r},t)$
produced by the $|e\rangle \rightarrow |b\rangle $ transition results in a
single excitation in the symmetrized atomic level $|b\rangle $. In the ideal
case, this coherently symmetrized state is \cite{duan01}
\begin{equation}
|\Phi _{1A}\rangle \sim \sum_{j=1}^{N}|a\rangle _{1}\ldots |b\rangle
_{j}\ldots |a\rangle _{N}.  \label{phi1}
\end{equation}

Although the initial detection of photon $1$ generated by the \textit{write }%
beam is probabilistic, the detection of photon $1$ results in the
conditional state $|\Phi _{1A}\rangle $ with one collective atomic
\textquotedblleft excitation\textquotedblright .\ This excitation can
subsequently be converted into an excitation of the light field with high
probability \textquotedblleft on demand\textquotedblright\ with a specified
emission direction and a programmable pulse shape \cite%
{duan01,vanenk98,duan02,fl}. In order to achieve the conversion from atoms
to field, a laser pulse from the \textit{read }beam tuned near the $%
|b\rangle \rightarrow |e\rangle $ transition illuminates the atomic sample,
thereby affecting the transfer $|b\rangle \rightarrow |a\rangle $ for the
sample with the accompanying emission of a second Raman photon $2$ on the $%
|e\rangle \rightarrow |a\rangle $ transition. For an optically thick atomic
sample, photon $2$ is emitted with high probability into a specified mode $%
\psi ^{(2)}(\vec{r},t)$ offset in time by $t^{(2)}=t^{(1)}+\delta t$. The
spatial and temporal structure of the modes $\psi ^{(1,2)}(\vec{r},t)$ are
set by the geometry of the atomic sample and by the shape and timing of the
\textit{write} and \textit{read} beams \cite{duan02}. In our experiment, the
modes of the (\textit{write}, \textit{read}) beams are spatially mode
matched, with measured visibility greater than $95$\% for the case of equal
frequency and polarization. The time delay $\delta t$ is limited in
principle only by the coherence time between the levels $|a\rangle $ and $%
|b\rangle $, which is long in practice.

The atomic sample for our experiment is provided by Cesium atoms in a
magneto-optical trap (MOT)\cite{metcalf99}, where the Cs hyperfine manifolds
$\{|6S_{1/2},F=4\rangle ,|6S_{1/2},F=3\rangle ,|6P_{3/2},F^{\prime
}=4\rangle \}$ correspond to the levels $\{|a\rangle ,|b\rangle ,|e\rangle
\} $, respectively. As illustrated by the timing diagram in Fig. \ref{layout}%
, the MOT is chopped from \textit{ON} to \textit{OFF} with $\Delta t=4\mu $%
s. In each cycle there is a \textquotedblleft dark\textquotedblright\ period
of duration $1\mu $s when all light responsible for trapping and cooling is
gated \textit{OFF}, with less than $0.1$\% of atoms measured to remain in
the $F=3$ level at this stage. The $j^{\text{th}}$ trial of the protocol for
single photon generation is initiated by a \textit{write} pulse which is
resonant with the $6S_{1/2},F=4\rightarrow 6P_{3/2},F^{\prime }=4$
transition at frequency $\omega _{4,4}$ and that has duration $\simeq 51$ ns
(FWHM). A critical parameter for the experiment is the resonant optical
thickness $\gamma _{4,4}$ of the atomic sample \cite{duan02}. We measure $%
\gamma _{4,4}\simeq 4-5$ for cw excitation, corresponding to an attenuation
of intensity $\exp (-\gamma _{4,4})$ in propagation through the MOT.

The \textit{write} pulse generates forward-scattered (anti-Stokes) Raman
light around frequency $\omega _{3,4}$ from the $F^{\prime }=4$ excited
level to the $F=3$ ground level ($|e\rangle \rightarrow |b\rangle $) that is
directed onto a single-photon detector \textit{D1}. After a variable delay $%
\delta t$, the \textit{read} pulse illuminates the sample, with this pulse
tuned to the $6S_{1/2},F=3\rightarrow 6P_{3/2},F^{\prime }=4$ transition at
frequency $\omega _{3,4}$ with duration $\simeq 34$ns (FWHM). Raman (Stokes)
light generated by the \textit{read} pulse around frequency $\omega _{4,4}$
from $F^{\prime }=4$ to $F=4$ ($|e\rangle \rightarrow |a\rangle $) is
directed onto a second single-photon detector \textit{D2}.

\begin{figure}[tb]
\includegraphics[width=8.6cm]{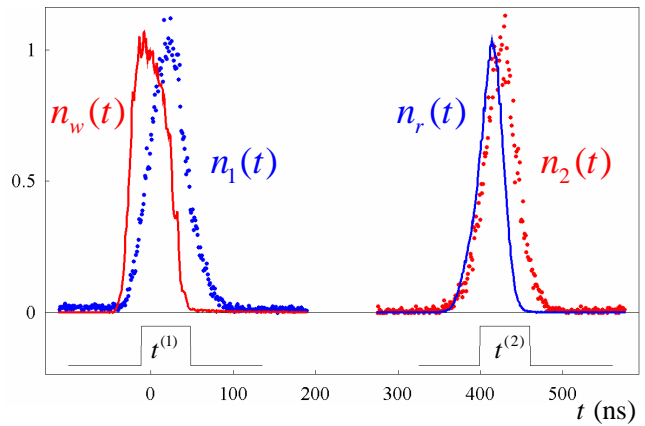}
\caption{Normalized singles counts $n_{i}(t)$ are shown for the \textit{write%
}, \textit{read}, and $(1,2)$ fields. The pulses around $t=0$ are from
detector \textit{D1} for the \textit{write} beam $n_{w}(t)$ (solid trace)
and for photon $1$, $n_{1}(t)$ (points). The pulses around $t=410$ns are
from detector \textit{D2} for the \textit{read} beam $n_{r}(t)$ (solid
trace) and for photon $2$, $n_{2}(t)$ (points). Note that in addition to the
symmetrized excitation, each \textit{write} pulse also transfers several
hundred atoms into the $F=3$ level due to spontaneous emission from its
near-resonant character. However, atoms transferred into $F=3$ via
spontaneous decay are spatially uncorrelated, so that their contribution to
the signal from the \textit{read} channel is strongly suppressed (by roughly
the fractional solid angle collected, $\protect\delta \Omega /4\protect\pi %
\simeq 4\times 10^{-5}$) as compared to the signal from single-atom
excitations of the form $|\Phi _{1A}\rangle $ \protect\cite{duan01}.}
\label{n1n2}
\end{figure}

By interchanging the frequencies for optical pumping of the filter cells
described in Fig. \ref{layout}, the (\textit{write}, \textit{read}) beams
can be detected at (\textit{D1}, \textit{D2}) in place of the $(1,2)$
fields. An example of the resulting pulse profiles accumulated over many
trials $\{j\}$ is presented in Fig. \ref{n1n2}, where the origin in time is
set to coincide with the approximate center of the \textit{write} pulse,
with the \textit{read} pulse following after a delay $\simeq 415$ns
determined by external control logic.

With the filter cells set to transmit the $(1,2)$ photons to the (\textit{D1}%
, \textit{D2}) detectors, respectively, we record histograms of the numbers $%
(n_{1}(t),n_{2}(t))$ of photoelectric events versus time, which are also
displayed in Fig. \ref{n1n2}. For the data presented here, the intensity of
the \textit{write} pulse is kept low ($\sim 10^{3}$ photons per pulse),
resulting in a time lag for the onset of the $n_{1}(t)$ counts in Fig. \ref%
{n1n2}. As discussed in the \textit{Supplemental Information}, the
probability $p_{write}^{(1)}$ to generate an anti-Stokes photon $1$ within
the solid angle of our imaging system is $p_{write}^{(1)}\simeq 10^{-2}$ per
pulse.

The \textit{read} pulse is about $100$ times more intense than the \textit{%
write }pulse, leading to high efficiency $\zeta _{3\rightarrow 4}\simeq 0.6$
for the transfer of population $|b\rangle \rightarrow |a\rangle $, with $%
p_{read}^{(2)}\simeq \zeta _{3\rightarrow 4}p_{write}^{(1)}$ for the Stokes
photon $2$. Examples of the resulting detection events $n_{2}(t)$ are shown
in Fig. \ref{n1n2}. In contrast to the behavior of $n_{1}(t)$, the intense
\textit{read }beam generates $n_{2}(2)$ counts promptly. More extensive
investigations of the timing characteristics of the emitted fields $(1,2)$
will be part of our subsequent investigations, including the relationship to
electromagnetically induced transparency (EIT) \cite{harris99,zibrov02}.

A virtue of the \textit{DLCZ} protocol is its insensitivity to a variety of
loss mechanisms, including inefficiencies in transport and detection of the $%
(1,2)$ photons. However, in an actual experiment, various non-ideal
characteristics of the atom-field interaction (as in our MOT) do lead to
deterioration of correlation for the $(1,2)$ photons (e.g., imperfect
filtering and/or background fluorescence as described in the caption of Fig. %
\ref{n1n2} and in the \textit{Supplemental Information}).
Fortunately there exists a well-defined border between the
classical and quantum domains for the $(1,2)$\ fields that can be
operationally accessed via coincidence detection, as was first
demonstrated in the pioneering work of Clauser \cite{clauser74}.

As illustrated in Fig. \ref{layout}b, electronic pulses from detectors (%
\textit{D1}, \textit{D2}) are\ separately gated with windows of duration $T=$
$60$ns centered on times $(t^{(1)},t^{(2)})$ corresponding to the
approximate peaks of the $(n_{1}(t),n_{2}(t))$ pulses shown in Fig. \ref%
{n1n2}. Photoelectric events that fall within the gate windows are directed
to a time-interval analyzer (TIA) configured in a standard fashion for
measurement of photoelectric correlations \cite{mandel-wolf-95}. For a
\textit{start} event from \textit{D1} within the interval $t_{j}^{(1)}\pm
T/2 $ for the $j^{\text{th}}$ trial of the experiment, the TIA records the
times of \textit{stop} events from \textit{D2} within successive intervals $%
t_{k}^{(2)}\pm T/2$. Over many repetitions of the experiment, we thereby
acquire time-resolved coincidences $n_{1,2}(\tau )$ between the $(1,2)$
fields, both within the same trial $k=j$ and for subsequent trials $%
k=j+1,j+2\ldots $ (i.e., a \textit{start} event from trial $j$ around time $%
t_{j}$ and a \textit{stop} event from trial $k$ around time $t_{k}$, where $%
t_{k}=t_{j}+(k-j)\Delta t$ for $k=j,j+1,\ldots $). By a $50$-$50\%$ beam
splitter, the field $1$ can be directed to detectors (\textit{D1}, \textit{D2%
}), and then in turn the field $2$ to (\textit{D1}, \textit{D2}). We thus
also acquire the time-resolved coincidences $n_{1,1}(\tau )$ and $%
n_{2,2}(\tau )$.

\begin{figure}[tb]
\includegraphics[width=8.6cm]{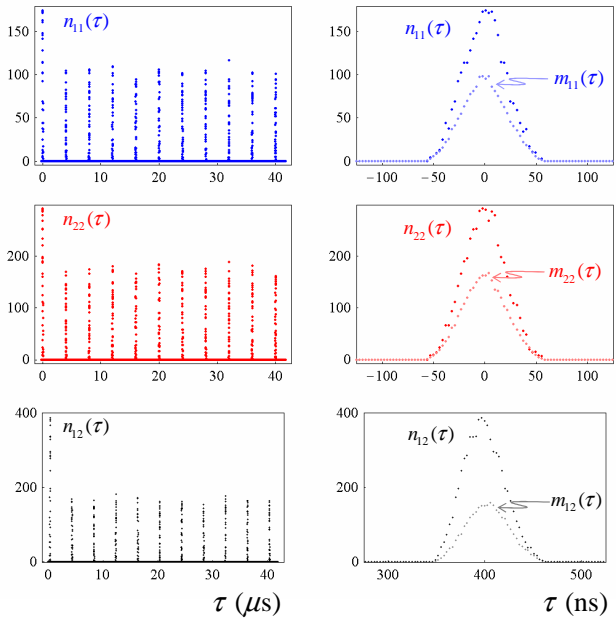}
\caption{Time-resolved coincidences $n_{\protect\alpha ,\protect\beta }(%
\protect\tau )$ between the $(1,1)$, $(2,2)$, and $(1,2)$ fields are
displayed versus time delay $\protect\tau $. \textit{Left column} - $n_{%
\protect\alpha ,\protect\beta }(\protect\tau )$ is shown over $11$
successive repetitions of the experiment. \textit{Right column -} The time
axis is expanded to a total duration of $250$ns with $\protect\tau =0$ set
to the center of the gating window $(t^{(1)},t^{(2)},t^{(1)})$ for $%
(n_{11},n_{22},n_{12})$, respectively. The larger peak $n_{\protect\alpha ,%
\protect\beta }(\protect\tau )$ corresponds to detection pairs from the same
trial $j$, while the smaller peak $m_{\protect\alpha ,\protect\beta }(%
\protect\tau )$ is for pairs from different trails as defined in the text.
Typical acquisition parameters are as follows. Detectors (\textit{D1},
\textit{D2}) have average count rates of about ($400$/s, $250$/s),
respectively, while background counts with no MOT present are about $100$/s.
Counts due to the MOT itself (with \textit{write} and \textit{read} beams
blocked) are less than ($10$/s, $20$/s) for (\textit{D1}, \textit{D2}). Dark
counts with the inputs to the fibers blocked are less than $5$/s. All these
numbers are for the gated-output mode of data acquisition as in Fig. \protect
\ref{layout} with $T=60$ns.}
\label{ncij}
\end{figure}

Figure \ref{ncij} displays an example of data accumulated in this manner for
coincidences $n_{\alpha ,\beta }(\tau )$ between the $(1,2)$, $(1,1)$, and $%
(2,2)$ beams, with successive peaks separated by the time between trials $%
\Delta t=4\mu $s. Note that there is an \textit{excess} of coincidence
counts in each of the initial peaks for joint detections from the same trial
$(\tau <\Delta t)$ as compared to $n_{\alpha ,\beta }(\tau )$ from different
trials $(\tau >\Delta t)$. This excess is shown more clearly in the plots in
the right column, which expand the time axis from the left column in Fig. %
\ref{ncij}. Here, data from successive trials $k=j+1,\ldots j+10$ have been
offset to $\tau <\Delta t$ and then averaged for comparison with $n_{\alpha
,\beta }(\tau )$ from the same trial $j$ by introducing the quantity $%
m_{\alpha ,\beta }(\tau )=\frac{1}{10}\sum_{k=j+1}^{j+10}n_{\alpha ,\beta
}(\tau +(k-j)\Delta t)$. As discussed in the \textit{Supplemental Information%
}, statistical independence for trials with $k\neq j$ is enforced by the
experimental protocol of reapplying the MOT\ and repuming beams after each
trial.

From the data in Fig. \ref{ncij}, we determine the total number of
coincidences $N_{\alpha ,\beta }=\sum_{\{\tau _{i}\}}n_{\alpha ,\beta }(\tau
_{i})$ with $(\alpha ,\beta )=(1,2)$ obtained by summing over time bins $%
\{\tau _{i}\}$ for detection within the same trial $j$, and $M_{\alpha
,\beta }=\sum_{\{\tau _{k}\}}m_{\alpha ,\beta }(\tau _{k})$ obtained from
\textit{start} and \textit{stop} events from different trials $(j\neq k)$.
Fields for which the Glauber-Sudarshan phase-space function is well-behaved
(i.e., \textit{classical} fields) are constrained by a Cauchy-Schwarz
inequality for the various coincidence counts [\textit{Supplemental
Information} and Ref. \cite{mandel-wolf-95}], namely
\begin{equation}
\left[ \tilde{g}_{1,2}(\delta t)\right] ^{2}\leq \tilde{g}_{1,1}\tilde{g}%
_{2,2}\text{ ,}  \label{cs}
\end{equation}%
where $\tilde{g}_{1,1}\equiv \frac{N_{1,1}}{M_{1,1}}$ , $\tilde{g}%
_{2,2}\equiv \frac{N_{2,2}}{M_{2,2}}$ , $\tilde{g}_{1,2}(\delta t)\equiv
\frac{N_{1,2}}{M_{1,2}}$.

For the data displayed in Fig. \ref{ncij}, we find $\tilde{g}%
_{1,1}=(1.739\pm 0.020)$ and $\tilde{g}_{2,2}=(1.710\pm 0.015)$, in
correspondence to the expectation that the $(1,2)$ fields should each
exhibit Gaussian statistics with $\tilde{g}_{1,1}=\tilde{g}_{2,2}=2$ for the
protocol of \textit{DLCZ }in the ideal case, but here degraded by diverse
sources of background counts (see \textit{Supplemental Information}). By
contrast, for the cross-correlations of the $(1,2)$ fields, we record $%
\tilde{g}_{1,2}(\delta t)=(2.335\pm 0.014)$, with $\delta t=405$ns. \textit{%
Hence the inequality of Eq. \ref{cs}\ for classical fields is strongly
violated, namely} $[\tilde{g}_{1,2}^{2}(\delta t)=5.45\pm 0.11]\nleq \lbrack
\tilde{g}_{1,1}\tilde{g}_{2,2}=2.97\pm 0.08]$, where all errors indicate the
statistical uncertainties. This violation of the Cauchy-Schwarz inequality
clearly demonstrates the nonclassical character of the correlations between
photons $(1,2)$ generated by the (\textit{write}, \textit{read}) beams.
Moreover, as discussed in more detail in the \textit{Supplemental Information%
}, the measured coincidence rates in Fig. \ref{ncij} explicitly document the
cooperative nature of the emission process. Overall, we estimate that the
probability $p_{c}^{(q)}$ for coincidence of the $(1,2)$ photons due to
collective atomic excitation as described by the state $|\Phi _{1A}\rangle $
is roughly $p_{c}^{(q)}\simeq 10^{-4}$ for each trial $j$, referenced to the
output of the MOT.

The temporal extent of the photon wave packet $\psi (\vec{r},t)$ for the $%
(1,2)$ photons is also of some interest. To investigate this issue, we have
carried out the experiment with expanded gate windows of duration $T=140$ns
that then encompass the entire domains over which counts $n_{1}(t)$ and $%
n_{2}(t)$ are observed in Fig. \ref{n1n2}. In this case, we record $\tilde{g}%
_{1,1}=(1.72\pm 0.04),\tilde{g}_{2,2}=(1.52\pm 0.05),$ and $\tilde{g}%
_{1,2}(\delta t)=(2.45\pm 0.10)$, now with $\delta t$ set to be $320$ns.
\textit{The classical inequality of Eq. \ref{cs} is once again not satisfied}%
; $[\tilde{g}_{1,2}^{2}(\delta t)=6.00\pm 0.50]\nleq \lbrack \tilde{g}_{1,1}%
\tilde{g}_{2,2}=2.61\pm 0.11]\text{ .}$ These results with $T=140$ns also
confirm that dead-time effects do not play a significant role in the current
experiment.

As described in the \textit{Supplemental Information}, the violation of the
Cauchy-Schwarz inequality of Eq. \ref{cs} in the ideal case can be much
larger than we have observed, namely $\left[ \tilde{g}_{1,2}(\delta t)\right]
^{2}/[\tilde{g}_{1,1}\tilde{g}_{2,2}]\simeq \left[ (1+p)/(2p)\right] ^{2}\gg
1$, where $p\ll 1$\ is the excitation probability. In our experiment, the
size of the violation of the inequality was limited mostly by uncorrelated
fluorescence from individual atoms in the atomic sample. This contribution
will be made smaller in future experiments by moving to off-resonant
excitation, which necessitates higher optical density. There is also a
significant limitation due the presence of the leakage light from the
\textit{read} pulse. This classical pulse is only $9$GHz away from the
single-photon field $2$ of interest, and is filtered by a factor exceeding $%
10^{-9}$. To achieve even stronger violation of the inequality, we must
further improve the filtering capability.

Our observations of nonclassical correlations between the $(1,2)$ photons
represent the first important step in the realization of the protocol of
\textit{DLCZ} \cite{duan01} for scalable quantum communication with atomic
ensembles, although it is not yet sufficient for realization of the full
protocol. Beyond the nonclassical correlations, our experiment also
demonstrates successful filtering of the various fields and collective
enhancement by the atomic ensemble, all of which are critical for
realization of the full quantum repeater protocol. More generally, the
capabilities that we have demonstrated should help to enable other advances
in the field of quantum information, including implementation of quantum
memory \cite{fl,schori} and fully controllable single-photon sources \cite%
{pelton02}, which, combined together, help to pave the avenue for
realization of universal quantum computation \cite{KLM}.

\textbf{Acknowledgements} HJK gratefully acknowledges interactions with M.
D. Lukin regarding various aspects of the experiment and its interpretation.
This work was supported by the National Science Foundation, by the Caltech
MURI Center for Quantum Networks under ARO Grant No. DAAD19-00-1-0374, and
by the Office of Naval Research.

\textbf{Competing interests statement} The authors declare that they have no
competing financial interests.

\textbf{Correspondence} and requests for materials should be addressed to
H.J.K. (e-mail: hjkimble@caltech.edu).

\newpage

\end{document}